\documentclass[preprint,12pt,authoryear]{elsarticle} 
\usepackage{amssymb} 
\usepackage{amsthm} 
\usepackage{lineno} 
\usepackage[utf8]{inputenc}
\usepackage{graphicx}
\usepackage{amsmath}
\usepackage{subfig}
\usepackage{float}
\journal{Journal of Theoretical Biology}

\begin{document}

\begin{frontmatter}
\title{Evolution of Cooperation in Public Goods Games with Stochastic Opting-Out}


\author[label1,label2]{Alexander G Ginsberg \corref{cor1}}
\ead{aaginsb@aol.com}
\cortext[cor1]{corresponding author}
\address[label1]{Department of Mathematics, Michigan State University, 619 Red Cedar Road, C212 Wells Hall, East Lansing, MI 48824, USA}
\address[label2]{Department of Mathematics, Dartmouth College, 27 N Main St., 6188 Kemeny Hall, Hanover, NH 03755, USA}
\author[label2,label3]{Feng Fu}
\ead{fufeng@gmail.com}
\address[label3]{Department of Biomedical Data Science, Geisel School of Medicine at Dartmouth, Lebanon, NH 03756, USA}

\begin{abstract}
This paper investigates the evolution of strategic play where players drawn from a finite well-mixed population are offered the opportunity to play in a public goods game.  All players accept the offer. However, due to the possibility of unforeseen circumstances, each player has a fixed probability of being unable to participate in the game, unlike similar models which assume voluntary participation. We first study how prescribed stochastic opting-out affects cooperation in finite populations. Moreover, in the model, cooperation is favored by natural selection over both neutral drift and defection if return on investment exceeds a threshold value defined solely by the population size, game size, and a player's probability of opting-out. Ultimately, increasing the probability that each player is unable to fulfill her promise of participating in the public goods game facilitates natural selection of cooperators. We also use adaptive dynamics to study the coevolution of cooperation and opting-out behavior. However, given rare mutations minutely different from the original population, an analysis based on adaptive dynamics suggests that the over time the population will tend towards complete defection and non-participation, and subsequently, from there, participating cooperators will stand a chance to emerge by neutral drift. Nevertheless, increasing the probability of non-participation decreases the rate at which the population tends towards defection when participating. Our work sheds light on understanding how stochastic opting-out emerges in the first place and its role in the evolution of cooperation.
\end{abstract}


\end{frontmatter}

\section{Introduction}
Cooperation is everywhere. (See Axelrod (1984), Hölldobler and Wilson (2009), Traulsen and Nowak (2006), and Trivers (1971)). Bacteria cooperate. For example, bacteria cooperate in biofilm production, where bacteria go so far as to use quorum sensing to determine when there are enough cooperators that contributing to the biofilm is worthwhile (Nadell 2008). Ants cooperate, building vast anthills where members of a colony live together. Birds cooperate, sounding an alarm when predators are nearby. Moreover, humans cooperate. Indeed, whenever we contribute to a joint hunting effort, bring food to a potluck, or work together to combat climate change, we are cooperating. Why, though, do we see cooperation in all walks of life? How does cooperation evolve? Researchers have dedicated significant effort in the past twenty years towards studying the evolution of cooperation. (See Antal et al. (2009), Boyd et al. (2010), Hauert et al. (2002a), Hauert et al. (2002b), Nowak (2006b), and Priklopil et al. (2017), as examples).
\\\\
In particular, one common type of social interaction in which cooperation frequently arises and which has recently attracted attention by researchers is the public goods game (PGG). (See Hauert et al. (2002a), Hauert et al. (2002b), and Pacheco et al. (2015)). In a public goods game, cooperators contribute to a common pool which all participants of the game then share equally. In fact, in all of the instances of cooperation mentioned in the preceding paragraph, organisms contribute to a public good. In the case of bacteria, the public good is biofilm production. For ants, the good is the anthill. For birds, the good is the knowledge that a predator is nearby and hence that they should be careful. Lastly, for the party-goers, the good is the food at the potluck.
\\\\
However, whenever cooperators contribute to a common pool, there are free-riders, who benefit from the common pool without contributing. Game theorists frequently refer to such free-riders as \textit{defectors}. These defectors cause the participants of the game to receive a smaller share of the common pool--a smaller payoff--than the social optimum where every player cooperates. In fact, regardless of what each other player does, a defector always earns a larger payoff because the defector does not have to contribute to the common pool, making defection the \textit{dominant strategy}. Game theorists refer to a situation in which the dominant strategy is not socially optimal as a \textit{social dilemma}. Consequentially, if each player were rational but unaware of the strategies of the other players, each player would choose to defect, and each player would receive no payoff. 
\\\\
In reality, even though in any particular PGG defectors will outperform cooperators, averaging over all games, it may be the case that cooperators actually outperform defectors. Such a situation is an example of Simpson's paradox (Hauert et al., 2002a). Additionally, there are many ways in which a tweak to the PGG may promote cooperation~\citep{Battiston_NJP17,Szolnoki_RSI15,Szolnoki_PRSB15}. For instance, kin selection (Antal et al., 2009)(Nowak, 2006b), punishment of defectors (Boyd et al., 2010), signaling (Pacheco et al. 2015), and optional participation (Hauert et al. (2002a) and Hauert et al. (2002b)), and combinations of the preceding methods (Sigmund et al. 2010)(Hauert et al., 2008) have been used to promote cooperation. However, in the literature, a small but realistic tweak to the public goods game has yet to be addressed. Specifically, even if there is no punishment of defectors or if players cannot opt-out, due to unforeseen circumstances, at times players simply cannot participate in the PGG. For instance, an individual traveling to a hunting party may come across a flooded road and be forced to turn back. Or, an individual going to an international conference on climate change may suddenly become too ill to travel. It is even possible that on a whim, an individual may decide to engage in some activity other than the game. As a result, players participate in the public goods game stochastically, unable to participate independently of whether or not the player plans to cooperate or defect.
\\\\
We investigate such public goods games with stochastic non-participation. Ultimately, we add a fully analyzed stochastic model to the literature, improving the understanding of the evolution of cooperation. Moreover, our model demonstrates that a tweak even more slight than others in the current literature, can facilitate cooperation. We conclude with an analysis of adaptive dynamics for simplified 2 person PGGs in finite populations. In such an analysis, we demonstrate that increasing the probability of non-participation temporarily slows the rate at which the population tends to defection when participating given rare mutations only minutely different from the original population.

\section{The Model}
We consider a well-mixed finite population of $n$ individuals, and suppose that frequently $N\leq n$ randomly selected individuals receive the opportunity to participate in a public goods game (PGG). In the PGG, individuals can choose to cooperate, investing 1 unit into a common pool, as in Hauert et al.(2002a) and Hauert et al. (2002b). Some force then multiplies the 1 unit each cooperator invests by some factor $N>r>1$ and thus for each unit invested by a cooperator, the common pool increases by $r$ units. At the end of the game, each PGG participant obtains an equal share of the common pool. However, the individuals who do not choose to cooperate choose to defect, receiving a share from the common pool without contributing. To simplify the model, we assume that individuals determine their strategies before the PGG has begun, ignoring group composition, as in Hauert et al. (2002a) and Hauert et al. (2002b).
\\\\
As stated, the preceding model leads to domination by defectors for all games where the multiplier $r$ is smaller than the game size $N$ and the game is thus a social dilemma. To promote cooperation, we assume that due to unforeseen circumstances each player has a fixed probability $\alpha$ of being unable to participate in the PGG, instead obtaining a fixed benefit or loss $\sigma$. 
\\\\
Furthermore, our model needs a method by which the population may change its composition of players cooperating or defecting. We take pairwise comparison as such a method, where occasionally two individuals are randomly selected. One individual will update his or her strategy by comparing his or her success to the other individual. We let the probability $p$ that the updating individual adopts the strategy of the other individual be proportional to the expected payoff difference between individuals of the two strategies. Specifically, we let the probability of changing strategies be given by the Fermi function, as in Traulsen (2007) and Pacheco (2015):
\begin{equation}
p=(1+\exp[-\gamma(\pi_{com}-\pi_{up})])^{-1}, 
\end{equation}
where $\pi_{com}$ represents the expected payoff of individuals playing the strategy of the individual selected for comparison, $\pi_{up}$ represents the expected payoff of individuals playing the strategy of the individual selected for updating, and $\gamma\geq 0$ represents a selection pressure, and corresponds to an inverse temperature (Traulsen et al., 2007). 
\begin{figure}[h!]
\label{Model Schematic of Stochastic Opting-out}
\includegraphics[width=\textwidth]{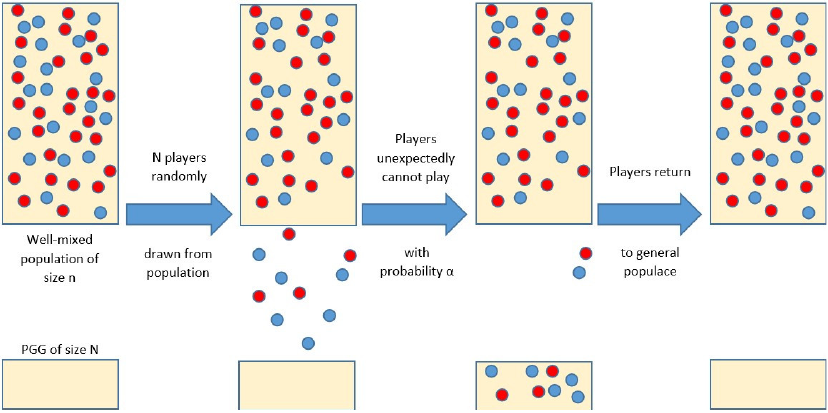}
\caption{Model Schematic of Stochastic Opting-out. Cooperators (blue) and defectors (red) are represented by dots. A fixed number of players are randomly drawn from the population to participate in a PGG, represented by the small tan rectangular area. While most players are able to make it to the game, some are not. Players then return to the general populace, where no game is occurring.}
\end{figure}
\\\\
When it comes to Adaptive Dynamics in finite populations, for simplicity, we assume game size is 2. Furthermore, applying adaptive dynamics to the problem as done in Imhof and Nowak (2010), we assume that a single mutant who plays a strategy similar to that to the original population invades the original population. Specifically, we suppose that every player plays a strategy in the strategy space $(\beta,\alpha)$, where $\beta$ is the probability that the player cooperates if he or she plays, and $\alpha$ is the probability that due to unforeseen circumstances the player cannot play. We let the original population be composed solely of players with strategy $(\beta,\alpha)$, and we suppose that the population is invaded by a single player with strategy $(\beta_1, \alpha_1)$. Then, we let $\beta_1 \rightarrow \beta$ and $\alpha_1 \rightarrow \alpha$. As in Imhof and Nowak (2010), we also assume rare mutation. That is, we assume sufficient time passes between mutations that either fixation, or extinction, of the mutant type occurs.

\section{Results}
\subsection{Pairwise Invasion Based on Fixation Probability}
To proceed with the analysis of the stochastic model, we must calculate the expected payoffs for cooperators and defectors, $\pi_c$ and $\pi_d$, respectively. To calculate $\pi_d$, we use the method presented by Hauert et al. (2002b). First, we observe that in a game with $S$ players, defectors receive a benefit $rn_c/S$, where $n_c$ is the number of cooperators in the game, if $S>1$. However, if $S=1$, that player must be a loner, and will receive payoff $\sigma$. Then, noting that any player does not play with probability $\alpha$ and plays with probability $1-\alpha$, and letting $x_c$ be the proportion of cooperators in the population,
\begin{equation}
\pi_d=\alpha\sigma+(1-\alpha)[rx_c[1-(1-\alpha^N)/(1-\alpha)]+\alpha^{N-1}\sigma].
\end{equation}
We defer the details to Appendix A.
Employing a similar method,
\begin{equation}
\pi_c=\pi_d+r/(n-1)[\alpha(1-\alpha^{N-1})]+(1-\alpha)[-1+(1-r)\alpha^{N-1}+(r/N)(1-\alpha^N)/(1-\alpha)].
\end{equation}
We defer the details to Appendix B.
Hence, 
\begin{equation}
\pi_c-\pi_d=r/(n-1)[\alpha(1-\alpha^{N-1})]+(1-\alpha)[-1+(1-r)\alpha^{N-1}+(r/N)(1-\alpha^N)/(1-\alpha)],
\end{equation}
a constant.
Then, inputting $\pi_c-\pi_d$ into (1), the probability that a cooperator becomes a defector given that a cooperator is selected for updating and a defector is selected for comparison is
\begin{equation}
p_{cd}=(1+\exp[\gamma(r/(n-1)[\alpha(1-\alpha^{N-1})]+(1-\alpha)[-1+(1-r)\alpha^{N-1}+(r/N)(1-\alpha^N)/(1-\alpha)])])^{-1},
\end{equation}
which is constant regardless of the number of cooperators. Thus the probability that the number of cooperators decreases by one in one iteration of the pairwise comparison model is \begin{equation}
p_{cd}i(N-i)/[N(N-1)].
\end{equation}
Likewise, the probability that a defector becomes a cooperator given that the defector is selected for updating and the cooperator is selected for comparison is 
\begin{equation}
p_{dc}=(1+\exp[-\gamma(r/(n-1)[\alpha(1-\alpha^{N-1})]+(1-\alpha)[-1+(1-r)\alpha^{N-1}+(r/N)(1-\alpha^N)/(1-\alpha)])])^{-1},
\end{equation}
also a constant. Hence, the probability that the number of cooperators increases by one in one iteration of the pairwise comparison model is
\begin{equation}
p_{dc}i(N-i)/[N(N-1)].
\end{equation}
Of course, though, if the number of cooperators, $i$, is 0 or n, the probabilities that a cooperator will change to a defector and that a defector will change to a cooperator are both zero, and the number of cooperators remains at 0 or $n$. That is, $i=0$ and $i=n$ are absorption states in the model.
\\\\
Moreover, now knowing $p_{cd}$ and $p_{dc}$, and noting that $p_{cd}+p_{dc}=1$, we can calculate the transition matrix $P$ for the Markov chain in which pairwise selection is iterated repeatedly. However, as the transition matrix itself is not vital for our analysis, we defer discussion of the transition matrix to Appendix C. On the other hand, the fixation probability of cooperation, that is, the probability that given i cooperators in a population of defectors that every individual will become a cooperator, \textit{is} vital. Following the procedure outlined by Nowak (2006a), we demonstrate that the fixation probability of cooperation given $i\geq 1$ cooperators, $x_i$, is
\begin{equation}
x_i=(1+\Sigma_{j=1}^{i-1} \Pi_{k=1}^jp_{cd}/p_{dc})/(1+\Sigma_{j=1}^{n-1} \Pi_{k=1}^jp_{cd}/p_{dc}),
\end{equation}
where $i=1$ implies the numerator is 1, where we denote $p_{cd}/p_{dc}$ by \\$G(\alpha,\gamma,N,n,r)$, and
\begin{equation}
G(\alpha,\gamma,N,n,r)=(1 +\exp(-\gamma(\pi_c-\pi_d)))/(1 + \exp(\gamma(\pi_c-\pi_d))).
\end{equation}
Notably, G is constant over $i$. Hence, we may expand the numerator and denominator of $x_i$ as geometric series. So, if $G\neq 1$, 
\begin{equation}
x_i=(1-G^i)/(1-G^n).
\end{equation}
However, $G=1$ implies that $p_{cd}=p_{dc}=1/2$, which implies
neutral drift. We assume for now that $G\neq 1$. Then, observing that $p_{dc}/p_{cd}=G^{-1}$, the fixation probability of defection given i defectors is simply $x_i$ with $G$ replaced by $G^{-1}$:
\begin{equation}
y_i=[G^{n-i}-G^n]/[1-G^n].
\end{equation}
Hence, the fixation probability given $i$ cooperators is 
\begin{equation}
y_{n-i}=[G^i-G^n]/[1-G^n].
\end{equation}
Thus, the probability of fixation of cooperators or defectors given i defectors is
\begin{equation}
x_i+y_{n-i}=1.
\end{equation}
Consequentially, the system always reaches an absorption state.
\\\\
Furthermore, now knowing the probabilities of fixation of cooperation given $i$ cooperators, $x_i$, and of defection given $i$ defectors, $y_i$, we can calculate the strategy favored by natural selection. Moreover, as in Nowak (2006a), natural selection favors cooperation over defection if and only if $x_1>y_1$. Likewise, natural selection favors defection over cooperation if and only if $y_1>x_1$ (Nowak, 2006a). Additionally, natural selection favors cooperation over neutral drift if and only if $x_1>1/n$ = the probability of fixation given natural drift (Nowak, 2006a). Likewise, natural selection favors defection over neutral drift if and only if $y_1>1/n$ (Nowak, 2006a). In fact, 
\begin{equation}
x_1>1/n \Leftrightarrow G<1.
\end{equation}
We defer the proof to appendix D. Also, $G=1$, implies neutral drift, since $G=1$ implies $p_{cd}=p_{dc}=1/2$. Since $G\neq1$ implies either $p_{cd}>p_{dc}$ or vice-versa, there is neutral drift if and only if $G=1$. Thus, $x_1<1/n$ if and only if $G>1$. 
Hence, natural selection favors cooperation over neutral drift if and only if $G<1$, favors neither cooperation nor neutral drift one over the other if and only if $G=1$, and favors neutral drift over cooperation if and only if $G>1$.
On the other hand, 
\begin{equation}
G<1 \Rightarrow y_1<1/n,
\end{equation} 
and
\begin{equation}
G>1 \Rightarrow y_1>1/n.
\end{equation}
We defer proofs of the two preceding assertions to Appendix D. Additionally if $G=1$, then there is neutral drift, as we demonstrated above, so $y_1=1/n$. Thus, if $G>1$, $y_1>1/n>x_1$; if $G=1$, then $y_1=1/n=x_1$; and otherwise, i.e $0<G<1$, $y_1<1/n<x_1$.
Additionally, noting that if the model is not experiencing neutral drift, then recalling that $(1 +\exp(-\gamma(\pi_c-\pi_d)))/(1 + \exp(\gamma(\pi_c-\pi_d)))$, clearly
\begin{equation}
G>1 \Leftrightarrow \pi_c-\pi_d<0.
\end{equation}
Likewise,
\begin{equation}
G<1 \Leftrightarrow \pi_c-\pi_d>0.
\end{equation}
Also, $G=1$ if and only if $\pi_c-\pi_d=0$. Thus, there are three possibilities:
\\\\
1) Natural selection favors cooperation over neutral drift, and neutral drift over defection ($\pi_c-\pi_d>0$), or
\\\\
2) Natural selection favors neither cooperation nor neutral nor defection one over the other, ($\pi_c-\pi_d=0$), or 
\\\\
3) Natural selection favors defection over neutral drift, and neutral drift over cooperation ($\pi_c-\pi_d<0$).
\\\\
Thus, the sign of $\pi_c-\pi_d$, a function of the probability that a given player opts out $\alpha$, the game size $N$, the population size $n$, and the return on investments by cooperators, $r$, exclusively determines which strategies, cooperation or defection, natural selection favors one over the another and whether or not natural selection favors each strategy over neutral drift. 
\begin{figure}
\subfloat[$\protect$]
{\includegraphics[width=0.5\textwidth]{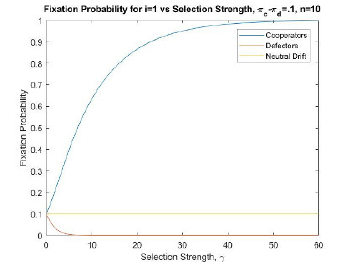}\label{fig:f2}}
  \hfill
  \subfloat[$\protect$]
{\includegraphics[width=0.5\textwidth]{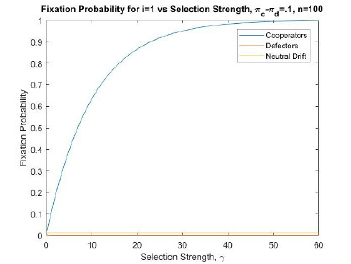}\label{fig:f3}}
  \hfill
  \subfloat[$\protect$]
{\includegraphics[width=0.5\textwidth]{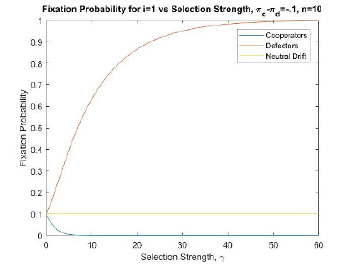}\label{fig:f4}}
\hfill
  \subfloat[$\protect$]{\includegraphics[width=0.5\textwidth]{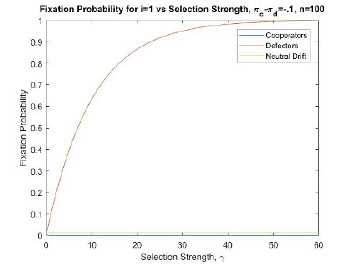}\label{fig:f5}}
\caption{Pairwise Invasion Dynamics in Finite Populations. Shown are graphs of fixation probabilities for $\pi_c-\pi_d>0$, as in (a) and (b), and for $\pi_c-\pi_d$, as in (c) and (d). If $\pi_c-\pi_d>0$, the fixation probability starting with one cooperator is always larger than the same given neutral drift which is in turn always larger than that of one defector. On the other hand, if $\pi_c-\pi_d<0$, then the situation is reversed. That is, the fixation probability starting with one \textit{defector} is always larger than the same given neutral drift which is in turn always larger than that of one \textit{cooperator}. Notably, the graphs for $\pi_c-\pi_d<0$, may be obtained from the graphs for $\pi_c-\pi_d>0$ simply by relabeling cooperators as defectors and vice-versa. This is because reversing the sign of $\pi_c-\pi_d$ is equivalent to inverting $p_{cd}/p_{dc}$. 
Also, the values chosen for $\pi_c-\pi_d$ are possible for the given values of n.}
\end{figure}
\\\\
This has profound implications. Primarily, there exists a minimum value of $r$, $R$, for given $N$ and $\alpha$ such that $r>R$ implies that $\pi_c-\pi_d>0$, $r<R$ implies that $\pi_c-\pi_d<0$, and $r=R$ implies that $\pi_c-\pi_d=0$. Indeed, recalling that $\pi_c-\pi_d=r/(n-1)[\alpha(1-\alpha^{N-1})]+(1-\alpha)[-1+(1-r)\alpha^{N-1}+(r/N)(1-\alpha^N)/(1-\alpha)]$, and noting that by the lemma in Appendix D, $[1-\alpha^N]/[N(1-\alpha)]-\alpha^{N-1} >0$, and thus that $[\alpha(1-\alpha^{N-1})]/[(n-1)(1-\alpha)]+[1-\alpha^N]/[N(1-\alpha)]-\alpha^{N-1}>0$, it follows that 
\begin{align}
\pi_c-\pi_d>&0 \Leftrightarrow\\
r>&\dfrac{1-\alpha^{N-1}}{{[\alpha(1-\alpha^{N-1})]/[(n-1)(1-\alpha)]+[1-\alpha^N]/[N(1-\alpha)]-\alpha^{N-1}}}=R(\alpha),
\end{align}
where $R(\alpha)$ is defined for [0,1).
Simplifying, 
\begin{equation}
R(\alpha)=N\dfrac{1-\alpha-\alpha^{N-1}+\alpha^N}{1+\alpha N/(n-1)-\alpha^{N-1}N+\alpha^N(N-1-N/(n-1))},
\end{equation}
demonstrating that on [0,1), $R$ is  also the quotient of two polynomials of degree $N$, and hence is continuous. By analogous proofs, $\pi_c-\pi_d=0$ if and only if $r=R(\alpha)$ and $\pi_c-\pi_d<0$ if and only if $r<R(\alpha)$. Additionally, on [0,1),
\begin{equation}
R(\alpha)<N,
\end{equation}
(we defer the proof to Appendix D), so it is always possible to choose $r$ such that natural selection favors cooperation. Moreover, as proven in Appendix E, $R(\alpha)$ is strictly decreasing on [0,1). Thus, given investment $r=R$, there is a threshold $\alpha_0$ such that $\alpha>\alpha_0$ implies natural selection favors cooperation.
This threshold is analogous to the threshold on the proportion of individuals who choose to opt-opt suggested by Hauert et al. (2002b), which deals with an infinite rather than finite population and with planned rather than unplanned non-participation. Moreover, this threshold is the value of $\alpha$ satisfying $r=R(\alpha)$. Thus, as $\alpha$ increases, the requirements on $r$ such that natural selection favors cooperation become less and less stringent. In other words, increasing the probability for players to be unable to participate facilitates cooperation.

\subsection{Adaptive Dynamics in Finite Populations}
Considering that increasing the probability of non-participation facilitates cooperation, it may be surprising that the adaptive dynamics for the two player game discussed in \textit{The Model} indicates that natural selection will push individuals to always defect when participating or to never participate. To see why, we consider a population consisting of two types of players, type one and type two, defined by their strategies $(\beta_1, \alpha_1)$ and $(\beta_2,\alpha_2)$, respectively. Otherwise maintaining the notation used in section 3.1, the expected payoff for players of type 1 is
\begin{align}
\begin{split}
\pi_1=&\dfrac{n - i}{n - 1}(1 - \alpha_1)(1 - \alpha_2)(r\beta_2/2 + r\beta_1/2- \beta_1) + \dfrac{i - 1}{n - 1}(1 - \alpha_1)^2\beta_1(r - 1) +\\& \sigma\alpha_1 + \sigma(1 - \alpha_1)(\dfrac{n - i}{n - 1}\alpha_2 + \dfrac{i - 1}{n - 1}\alpha_1). 
\end{split}
\end{align}
We defer the derivation of the expected payoff for players of type 1 to Appendix F. Moreover, since the game is symmetric, the expected payoff for players of type 2 may be determined simply by replacing the number of players of type 1, i, with the number of players of type 2, n-i, and by switching subscripts. Specifically, the expected payoff for players of type 2 is
\begin{equation}
\begin{split}
\pi_2=&\dfrac{i}{n - 1}(1 - \alpha_2)(1 - \alpha_1)(r\beta_1/2 + r\beta_2/2- \beta_2) + \dfrac{n-i- 1}{n - 1}(1 - \alpha_2)^2\beta_2(r - 1)+\\& \sigma\alpha_2 + \sigma(1 - \alpha_2)(\dfrac{i}{n - 1}\alpha_1 + \dfrac{n-i - 1}{n - 1}\alpha_2). 
\end{split}
\end{equation}
Continuing to use the pairwise comparison model, the probability that a player of type 1 will adopt the strategy of a player of type 2 given that the player of type 1 updates and the player of type 2 compares is 
\begin{equation}
p_{1\rightarrow 2}=(1+\exp[-\gamma(\pi_2-\pi_1)])^{-1},
\end{equation}
where $\gamma$ is a selection pressure, just as in section 3.1. Similarly, the analogous probability for players of type  2 is 
\begin{equation}
p_{2\rightarrow 1}=(1+\exp[-\gamma(\pi_1-\pi_2)])^{-1}.
\end{equation}
Then, again following the method proposed by Nowak (2006a), the fixation probability of a player of type 1 given i players of type 1 in a population of players of type 2 is
\begin{equation}
x_i=(1+\Sigma_{j=1}^{i-1} \Pi_{k=1}^jp_{1\rightarrow 2}/p_{2 \rightarrow 1})/(1+\Sigma_{j=1}^{n-1} \Pi_{k=1}^jp_{1\rightarrow 2}/p_{2\rightarrow 1}).
\end{equation}
For the remainder of this section, we will assume players of type 2 compose the invaded population, and hence we will drop the subscripts on $\alpha_2$ and $\beta_2$.
\\\\
To investigate the adaptive dynamics, consider 
\begin{center}
$\vec{f}(\beta,\alpha)=\lim_{(\alpha_1,\beta_1)\rightarrow(\alpha,\beta)}(\partial x_1/\partial \alpha_1,\partial x_1/\partial \beta_1)$.
\end{center}
The direction given by $\vec{f}$ for $(\alpha,\beta)$, plotting f as a vector field, is the direction in the strategy space which maximizes the fixation probability of the invading mutant population given one invading mutant, $x_1$. Following the directions which maximize $x_1$ in the strategy space starting at an initial $(\alpha,\beta)$, that is, following the streamlines of $\vec{f}$, indicates the most likely path in the strategy space that a population will take as mutants with similar strategies eventually fixate in the population, as suggested by Imhof and Nowak (2010). Moreover, applying the StreamPlot function of Mathematica to the model for various combinations of $r$ and $\sigma$ in a population of size $n$ indicates that the probability an individual cooperates will decrease and that increasing $r$ or decreasing $\sigma$ will facilitate participation. 
Notably, Mathematica demonstrates that for $\gamma=1$
\begin{equation}
\lim_{(\alpha_1,\beta_1)\rightarrow(\alpha,\beta)}(\partial x_1/\partial \alpha_1)=(\alpha-1)(n-2)((r-1)\beta-\sigma)/(2n), 
\end{equation}
and
\begin{equation}
\lim_{(\alpha_1,\beta_1)\rightarrow(\alpha,\beta)}(\partial x_1/\partial \beta_1)=(1-\alpha)^2(2-2n-2r+nr)/(4n).
\end{equation}
Unfortunately, the problem proves too complicated to calculate a closed-form solution of $\vec{f}$ for every $\gamma$. Nevertheless, we conjecture that for any $\gamma$,
\begin{equation}
\lim_{(\alpha_1,\beta_1)\rightarrow(\alpha,\beta)}(\partial x_1/\partial \alpha_1)=(\alpha-1)\gamma(n-2)((r-1)\beta-\sigma)/(2n),
\end{equation}
and
\begin{equation}
\lim_{(\alpha_1,\beta_1)\rightarrow(\alpha,\beta)}(\partial x_1/\partial \beta_1)=(1-\alpha)^2\gamma(2-2n-2r+nr)/(4n).
\end{equation}
(30) and (31) imply (28) and (29), respectively, and hold for a variety of other test values of $n$, $r$, $\sigma$, and $\gamma$. In particular, (30) and (31) both hold if $\beta=0$.
\begin{figure}
\includegraphics[width=\textwidth]{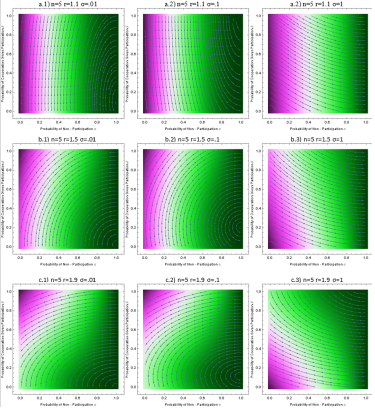}\label{}
\caption{Coevolution of Cooperation and Stochastic Opting-out. Shown are the adaptive dynamics using the stream plot function of Mathematica in a finite population of size $n = 5$ for selection strength $\gamma=1$ and various values of return on investment, $r$, and payoff for non-participants $\sigma$. Following the arrows leads to the most likely path the population will take in the strategy space. Observe that as $r$ increases or $\sigma$ decreases the population tends to participation at a faster rate or the population tends to non-participation at a slower rate, where the rate is indicated by the slope of the arrows. Also, the population always tends to defection when participating, and otherwise the population appears to tend to non-participation. However, increasing $\alpha$ decreases the rate at which individuals in the population tend to complete defection.} 
\end{figure}
\\\\
Furthermore, if (31) is indeed true, then $\lim_{(\alpha_1,\beta_1)\rightarrow(\alpha,\beta)}(\partial x_1/\partial \beta_1)$ would be independent of $\beta$ and $\sigma$.
Moreover, if $n>2$, $\alpha_2<1$, $\gamma>0$, and (31) is valid, simple algebraic manipulation yields
\begin{equation}
\lim_{(\alpha_1,\beta_1)\rightarrow(\alpha,\beta)}(\partial x_1/\partial \beta_1)>0 \Leftrightarrow r>1+n/(n-2)>2.
\end{equation}
Likewise, $\lim_{(\alpha_1,\beta_1)\rightarrow(\alpha,\beta)}(\partial x_1/\partial \beta_1)=0$ if and only if $r=1+n/(n-2)$, and $\lim_{(\alpha_1,\beta_1)\rightarrow(\alpha,\beta)}(\partial x_1/\partial \beta_1)<0$ if and only if $r<1+n/(n-2)$.
On the other hand, if $\gamma=0$, or $\alpha_2=1$, then $\lim_{(\alpha_1,\beta_1)\rightarrow(\alpha,\beta)}(\partial x_1/\partial \beta_1)=0$, and if $n=2$, $\lim_{(\alpha_1,\beta_1)\rightarrow(\alpha,\beta)}(\partial x_1/\partial \beta_1)=(-1/4)(1-\alpha)^2\gamma$. Hence, for all games which are social dilemmas, i.e $r<2$, and even for some games which are not social dilemmas, the presence of rare and minute mutations leads each individual towards defection when participating as long as $\alpha<1$ and $\gamma>0$.
\\\\
Moreover, if (31) holds, then cooperation is never stable if initially $\alpha \nrightarrow 1$, and each individual in the population tends towards always defecting, i.e, $\beta\rightarrow 0$. Thus, the sign of the right-hand-side of (30) becomes the sign of $\sigma$ when $n>2$ and 0 otherwise, (although if $n=2$ the right-hand side of (30) is \textit{always} 0).
Hence, if $\sigma<0$, each individual in the population tends towards always defecting and always participating if (30) is valid. If $\sigma=0$ or $n=0$, (30) and (31) demonstrate that defection with some degree of participation is stable, but cooperation is not. Instead, if $\sigma>0$, (30) and (31) show that $\alpha \rightarrow 1$ anyways. On the other hand, if initially $\alpha \rightarrow 1$, $\alpha$ remains near $1$ by (30) and (31). However, if $\alpha \rightarrow 1$, nobody participates (everyone gets the same payoff $\sigma$) and thus neutral drift allows the establishment of cooperation along the edge $\alpha =1$. Also, if (31) is valid, increasing $\alpha$ increases $\lim_{(\alpha_1,\beta_1)\rightarrow(\alpha,\beta)}(\partial x_1/\partial \beta_1)$, i.e increasing $\alpha$ decreases the rate at which individuals in the population tend towards complete defection. Moreover, from Eq. (31), we can obtain a possible rest point $\beta^* = \sigma/(r-1)$ on the edge of $\alpha =1$, as long as the value of $\beta^*\in (0,1)$. This can be confirmed in Fig.~3(b2) and Fig.~3(c2).

\section{Conclusion}
Notably, for games where every player participates, i.e $\alpha=0
$, the threshold return on investment, $R(0)$, above which natural selection favors cooperation is the game size, $N$. Hence, if the game size is reduced by a factor $(1-\alpha)$, where $0<\alpha<1$, the threshold value on investment is $N(1-\alpha)$. Moreover, if instead $\alpha$ is the probability that any given player does not participate, the law of large numbers suggests that for very large game sizes, the number of people participating in the game will be $N(1-\alpha)$. If the game size is very large but is small with respect to the population size, this is exactly the threshold value $R(\alpha)$ for return on investment above which natural selection favors cooperation. Hence, for very large games which are small with respect to the population size, reducing $\alpha$ appears to be the sole factor which facilitates cooperation. 
\begin{figure}[H]
\subfloat[$\protect$]
{\includegraphics[width=.5\textwidth]{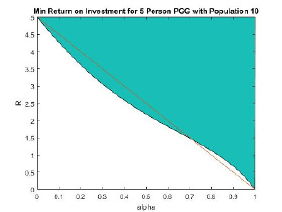}\label{fig:f7}}
  \hfill
  \subfloat[$\protect$]
{\includegraphics[width=.5\textwidth]{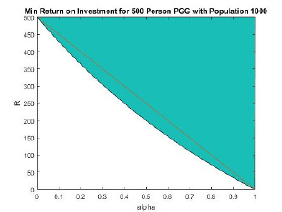}\label{fig:f8}}
\hfill
  \subfloat[$\protect$]
{\includegraphics[width=.5\textwidth]{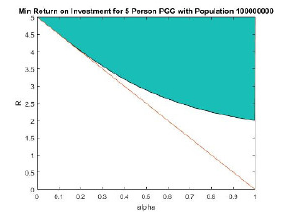}\label{fig:f9}}  \subfloat[$\protect$]
{\includegraphics[width=.5\textwidth]{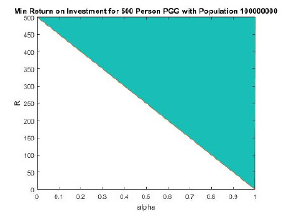}\label{fig:f10}}
  \hfill
\caption{The shaded areas represent combinations of the multiplier $r$ and probability of non-participation $\alpha$ which promote cooperation for game size $N=5$ and $n=10$ in (a), $N=500$ and $n=1000$ in (b), $N=5$ and $n=100000000$ in (c), and $N=500$ and $n=100000000$ in (d). The red line in each figure is the curve $N(1-\alpha)$. Note that in (b), $R(\alpha)\approx N(1-\alpha)/(1+1/2\alpha)<N(1-\alpha)$,  whereas in (c), $R\approx N\dfrac{1-\alpha-\alpha^{N-1}+\alpha^N}{1-\alpha^{N-1}N+\alpha^N(N-1)}>N(1-\alpha)$ for $\alpha>0$, which inequality may be obtained easily by noting that the denominator is positive (see the lemma in Appendix D), multiplying both sides by the denominator, and then applying simple algebraic manipulation. Also note that for $n>>N>>0$ as in (d), $R\approx N(1-\alpha)$. The approximation in (c) is obtained easily by taking the appropriate limits. The approximations in (b) and (d) are justified in Appendix G.} 
\end{figure}
\noindent
However, if population size and game size are both very large but population size is no longer arbitrarily large with respect to game size, there is a second factor at work. In the equation for the threshold, $R(\alpha)$, this second factor arises from the term $((n-1)/N)\alpha(1-\alpha^{N-1})/[1-\alpha]$. This factor, always positive, reduces the threshold R at every positive value of $\alpha$, thereby facilitating cooperation. Furthermore, if the game size is small but population size is still very large population, $R \rightarrow 2$ as $\alpha\rightarrow 1$. Also considering that $R$ is strictly decreasing implies that the threshold curve $R(\alpha)$ also satisfies $R(\alpha)>N(1-\alpha)$ for large $\alpha$, so there appears to be some other factor which resists cooperation in small groups. Specifically, this factor occurs at least in part because for $0<<\alpha<1$, games become rare and the vast majority of games become two player games, where natural selection favors cooperation if and only if the return on investment by cooperators is larger than 2. 
\\\\
Despite increasing $\alpha$ always reducing the threshold value for cooperation, the adaptive dynamics suggest that in the presence of minute and rare mutations individuals in the population always tend towards always defecting or never participating. Nevertheless, the adaptive dynamics also indicate that the rate at which the population tends to defection is slower for larger values of $\alpha$. So, while assuming that players have a fixed probability of non-participation in the adaptive dynamics does not make cooperation stable in the presence of rare and minute mutations, by decreasing the rate at which the population tends to defection, it essentially increases the time which the population spends at higher levels of cooperation. Also, cooperation emerges on the edge $\alpha=1$. Ultimately, either in the presence of rare and minute mutations or not, assuming players are unable to participate with a fixed probability facilitates cooperation.

\section*{Acknowledgments}
The authors would like to thank the National Science Foundation and Dartmouth College for funding the REU program at which the research was conducted. In particular, Alexander G Ginsberg would also like to thank professor Feng Fu, professor Anne Gelb, Tracy Moloney, and Amy Powell, all of Dartmouth College, for personally overseeing the program. Lastly, he would like to thank professors Ignacio Uriarte-Tuero, George Pappas, Tsvetanka Tsendova, and Teena Gerhardt, all of Michigan State University, for making sure he attended a program that fit his needs. Feng Fu acknowledges generous support from the Dartmouth Startup Fund, the Walter and Constance Burke Research Initiation Award, the NIH (grant no. C16A12652-A10712), and DARPA (grant no. D17PC00002-002)
\pagebreak
\appendix
\section{Derivation of $\pi_d$}
We define the probability that an event E occurs be denoted by P(E), and let the probability that E occurs given a second event F occurs be denoted by $P(E|F)$. Then,
\begin{align*}
\pi_d=&\alpha\sigma+ \Sigma P(n_c \cap S \cap plays)*payoff\\ 
=&\alpha\sigma+ \Sigma P(plays)P(S|plays)P(n_c |S \cap plays)*payoff\\
=&\alpha\sigma+(1-\alpha)\Sigma P(S|plays)P(n_c |S \cap plays)*payoff\\
=&\alpha\sigma+(1-\alpha)[\Sigma_{S=2}^{N}P(S|plays)\Sigma_{n_c=0}^{S-1}P(n_c|S\cap plays)rn_c/S+P(S=1|plays)\sigma]\\ 
=&\alpha\sigma+(1-\alpha)[\Sigma_{S=2}^{N}P(S|plays)/S \Sigma_{n_c=0}^{S-1}P(n_c|S\cap plays)rn_c+P(S=1|plays)\sigma]\\ 
\begin{split}
=&\alpha\sigma+(1-\alpha)[r\Sigma_{S=2}^{N}(1/S){N-1 \choose S-1}(1-\alpha)^{S-1}\alpha^{N-S}\Sigma_{n_c=0}^{S-1}{S-1\choose n_c}x_c^{n_c}*\\&(1-x_c)^{S-1-n_c}n_c+\alpha^{N-1}\sigma]\\ 
=&\alpha\sigma+(1-\alpha)[r\Sigma_{S=2}^{N}{N-1 \choose S-1}(1-\alpha)^{S-1}\alpha^{N-S}x_c(S-1)/S\Sigma_{n_c=1}^{S-1}{S-2\choose n_c-1}x_c^{n_c-1}*\\&(1-x_c)^{S-1-n_c}+\alpha^{N-1}\sigma]\\
=&\alpha\sigma+(1-\alpha)[r\Sigma_{S=2}^{N}{N-1 \choose S-1}(1-\alpha)^{S-1}\alpha^{N-S}x_c(S-1)/S\Sigma_{k=0}^{S-2}{S-2\choose k}x_c^{k}*\\&(1-x_c)^{S-2-k}+ \alpha^{N-1}\sigma]\\
=&\alpha\sigma+(1-\alpha)[r\Sigma_{S=2}^{N}{N-1 \choose S-1}(1-\alpha)^{S-1}\alpha^{N-S}x_c(S-1)/S+\alpha^{N-1}\sigma]\\
=&\alpha\sigma+(1-\alpha)[rx_c(\Sigma_{S=2}^{N}{N-1 \choose S-1}(1-\alpha)^{S-1}\alpha^{N-S}- \Sigma_{S=2}^{N}{N-1 \choose S-1}(1-\alpha)^{S-1}*\\&\alpha^{N-S}/S)+\alpha^{N-1}\sigma]\\
=&\alpha\sigma+(1-\alpha)[rx_c(\Sigma_{k=1}^{N-1}{N-1 \choose k}(1-\alpha)^k\alpha^{N-k-1}-(1/N)\Sigma_{S=2}^{N}{N \choose S}(1-\alpha)^{S-1}*\\&\alpha^{N-S})+\alpha^{N-1}\sigma]\\
=&\alpha\sigma+(1-\alpha)[rx_c((1-\alpha^{N-1})-1/(N(1-\alpha))\Sigma_{S=2}^{N}{N \choose S}(1-\alpha)^{S}\alpha^{N-S})+\alpha^{N-1}\sigma]\\
=&\alpha\sigma+(1-\alpha)[rx_c((1-\alpha^{N-1})-[1-N(1-\alpha)\alpha^{N-1}-\alpha^N]/[N(1-\alpha)]+\alpha^{N-1}\sigma]\\
=&\alpha\sigma+(1-\alpha)[rx_c(N-N\alpha-[1-\alpha^N])/(N[1-\alpha])+\alpha^{N-1}\sigma]\\
=&\alpha\sigma+(1-\alpha)[rx_c[1-(1-\alpha^N)/(1-\alpha)]+\alpha^{N-1}\sigma]. 
\end{split}
\end{align*}

We have verified via Mathematica and via Hauert et al. (2002b) that 
\begin{equation*}
\begin{split}
&\Sigma_{S=2}^{N}P(S|plays)\Sigma_{n_c=0}^{N-1}P(n_c|S\cap plays)rn_c/S+P(S=1|plays)\sigma=\\& rx_c[1-(1-\alpha^N)/(1-\alpha)]+\alpha^{N-1}\sigma.
\end{split}
\end{equation*}

\section{Derivation of $\pi_c$}
\begin{align*}
\begin{split}
\pi_c=&\alpha\sigma+ \Sigma P(n_c \cap S \cap plays)*payoff\\ 
=&\alpha\sigma+(1-\alpha)\Sigma P(S|plays)P(n_c |S \cap plays)*payoff\\ 
=&\alpha\sigma+(1-\alpha)[\Sigma_{S=2}^{N}P(S|plays)\Sigma_{n_c=0}^{S-1}P(n_c|S\cap plays)((r/S)(n_c+1)-1)+\\&P(S=1|plays)\sigma]\\ 
=&\alpha\sigma+(1-\alpha)[\Sigma_{S=2}^{N}P(S|plays)\Sigma_{n_c=0}^{S-1}P(n_c|S\cap plays)rn_c/S+\\&\alpha^{N-1}\sigma+\Sigma_{S=2}^{N}P(S|plays)\Sigma_{n_c=0}^{S-1}P(n_c|S\cap plays)(r/S-1)]. 
\end{split}
\end{align*}
Noting that
\begin{equation*}
\alpha\sigma+(1-\alpha)[\Sigma_{S=2}^{N}P(S|plays)\Sigma_{n_c=0}^{S-1}P(n_c|S\cap plays)rn_c/S+\alpha^{N-1}\sigma]=\pi_d, 
\end{equation*}
where $x_c$ is replaced by $x_c-1/(n-1)$, it follows that 
\begin{align*}
\begin{split}
\pi_c=&\pi_d+r/(n-1)[\alpha(1-\alpha^{N-1})]+(1-\alpha)\Sigma_{S=2}^{N}P(S|plays)\Sigma_{n_c=0}^{S-1}P(n_c|S\cap plays)*\\&(r/S-1)\\
=&\pi_d+r/(n-1)[\alpha(1-\alpha^{N-1})]+(1-\alpha)\Sigma_{S=2}^{N}P(S|plays)(r/S-1)\\
=&\pi_d+r/(n-1)[\alpha(1-\alpha^{N-1})]+(1-\alpha)[r\Sigma_{S=2}^{N}(1/S){N-1 \choose S-1}(1-\alpha)^{S-1}\alpha^{N-S}-\\& \Sigma_{S=2}^{N}{N-1 \choose S-1}(1-\alpha)^{S-1}\alpha^{N-S}]\\
=&\pi_d+r/(n-1)[\alpha(1-\alpha^{N-1})]+(1-\alpha)[(r/N)\Sigma_{S=2}^{N}{N \choose S}(1-\alpha)^{S-1}\alpha^{N-S}-\\& \Sigma_{k=1}^{N-1}{N-1 \choose k}(1-\alpha)^k\alpha^{N-k-1}]\\
=&\pi_d+r/(n-1)[\alpha(1-\alpha^{N-1})]+(1-\alpha)[(r/[N(1-\alpha)])\Sigma_{S=2}^{N}{N \choose S}(1-\alpha)^S\alpha^{N-S}-\\&(1-\alpha^{N-1})]
\end{split}
\end{align*}
\begin{align*}
\begin{split}
=&\pi_d+r/(n-1)[\alpha(1-\alpha^{N-1})]+(1-\alpha)[r/[N(1-\alpha)](1-N(1-\alpha)\alpha^{N-1}-\alpha^N)-\\&1+\alpha^{N-1}]
\end{split}
\end{align*}
\begin{align*}
\begin{split}
=&\pi_d+r/(n-1)[\alpha(1-\alpha^{N-1})]+(1-\alpha)[-1-r\alpha^{N-1}+\alpha^{N-1}+(r/N)(1-\alpha^N)/\\&(1-\alpha)]
\end{split}
\end{align*}
\begin{align*}
\begin{split}
=&\pi_d+r/(n-1)[\alpha(1-\alpha^{N-1})]+(1-\alpha)[-1+(1-r)\alpha^{N-1}+(r/N)(1-\alpha^N)/\\&(1-\alpha)].
\end{split}
\end{align*}
Again, we have verified via Mathematica and via Hauert et al.  (2002b) that
\begin{align*}
&\Sigma_{S=2}^{N}P(S|plays)\Sigma_{n_c=0}^{S-1}P(n_c|S\cap plays)(r/S-1)=\\&-1+(1-r)\alpha^{N-1}+(r/N)(1-\alpha^N)/(1-\alpha).
\end{align*}

\section{Transition Matrix}
We define $P$ be the transition matrix for the Markov chain formed by repeatedly iterating pairwise comparison. Then, $P_{i,i-1}=p_{cd}i(n-i)/[n(n-1)]$, and $P_{i,i+1}=p_{dc}i(n-i)/[n(n-1)]$, for i=2, 3, ..., n-1. Since the only other transition from i cooperators per iteration is the absence of transition, $P_{i,i}=1-P_{i,i-1}-P_{i,i+1}$, and the remaining entries in the $i^{th}$ row are 0. Also considering that $i=0$ cooperators and $i=n$ cooperators are absorbing states, it follows that P is the tridiagonal (n+1)x(n+1) matrix 
\begin{equation*}
\begin{bmatrix}
1 & 0 & 0 & 0 & \hdots &0 \\
P_{2,1} & P_{2,2}& P_{2,3} & 0& \hdots & 0 \\
0 & P_{3,2}&P_{3,3} &P_{3,4} & \ddots & \vdots \\
\vdots & \ddots & \ddots & \ddots & \ddots &0\\
0 & \hdots& 0 & P_{n-1,n-2} & P_{n-1,n-1}& P_{n-1,n}\\
0 & \hdots & 0 & 0 & 0 & 1
\end{bmatrix}.
\end{equation*}
Fortunately, the calculation $P^k$ as $k\rightarrow\infty$ is relatively straightforward. Indeed, the calculated the fixation probabilities $x_i$ in (11), and $y_{n-i}$ in (13), represent, respectively, the last and first entries in the $i^{th}$ row of $\lim_{n\rightarrow\infty}P^n$. Also considering that the entries in any given row of $P^n$ must sum to 1 as $P^n$ is a stochastic matrix, and that $x_i+y_{n-i}=1$, it follows that
\begin{equation*}
\lim_{n\rightarrow\infty}P^n=
\begin{bmatrix}
1 & 0 & \hdots & 0 & 0\\
x_1 & 0 &\hdots & 0 & y_{n-1}\\
x_2 & 0 &\hdots & 0 & y_{n-2}\\
\vdots & \vdots & \vdots & \vdots & \vdots\\
x_{n-1} & 0 & \hdots & 0 & y_{1}\\
0 & 0 & \hdots & 0 & 1
\end{bmatrix}.
\end{equation*}
Thus, $\lim_{n\rightarrow\infty}XP^n$ converges to a vector of the form $(a, 0, ..., 0,b)$. Namely, the set of vectors of the form (a, 0, ..., 0, b) is the set of eigenvectors of $\lim_{n\rightarrow\infty}P^n$, which in turn is the set of eigenvectors of P with eigenvalue 1. Moreover, if $X=(Prob(i=0), Prob(i=1), ..., Prob(i=n))$, then $\lim_{n\rightarrow\infty}XP^n$ converges to a vector of the form $(\alpha, 0, ..., 0, \beta)$, where $\alpha+\beta=1$. Since the set of vectors of the form $(\alpha, 0, ..., 0, \beta)$ with $\alpha+\beta=1$ is the set of stochastic eigenvectors of $P$ with eigenvalue 1, it follows that depending on the initial probability vector for the system, $X=(Prob(i=0), Prob(i=1), ..., Prob(i=n))$, the system can potentially converge to any stochastic eigenvector.
\section{Inequalities}
\subsection{Proof of (15)}
\begin{align}
&x_1>1/n \Leftrightarrow \\
&[1-G]/[1-G^n]>1/n \Leftrightarrow \\
&[1-G^n]/[1-G]<n \Leftrightarrow \\
&\Sigma_{k=0}^{n-1}G^k<\Sigma_{k=0}^{n-1}1 \Leftrightarrow\\
&G<1. \square
\end{align}
\subsection{Proof of (16)}
\begin{align}
&y_1<1/n \Leftrightarrow\\
&[G^{n-1}-G^n]/[{1-G^n}]<1/n \Leftrightarrow\\
&[1-G]/[1-G^n]<1/(nG^{n-1}) \Leftrightarrow\\
&[1-G^n]/[1-G]>nG^{n-1} \Leftrightarrow\\
&(1/n)\Sigma_{k=0}^{n-1}G^k>G^{n-1} \Leftrightarrow\\
&(1/n)\Sigma_{k=0}^{n-1}G^k>(G^{(n-1)(n)/2})^{2/n} \Leftrightarrow \\
&(1/n)\Sigma_{k=0}^{n-1}G^k>((\Pi_{k=0}^{n-1}G^k)^{1/n})^2. 
\end{align}
Moreover, if $G<1$, then $(\Pi_{k=0}^{n-1}G^k)^{\dfrac{1}{n}}>((\Pi_{k=0}^{n-1}G^k)^{\dfrac{1}{n}})^2$. Hence, if $G<1$, applying the arithmetic-mean-geometric-mean (AM-GM) inequality demonstrates that 
\begin{equation}
(1/n)\Sigma_{k=0}^{n-1}G^k>((\Pi_{k=0}^{n-1}G^k)^{1/n})^2.
\end{equation}
Thus, $G<1$ implies that $y_1<1/n$. $\square$
\subsection{Proof of (17)}
If $G>1$ and $n>1$, note that
\begin{align}
&y_1>1/n \Leftrightarrow\\
&[G^{n-1}-G^n]/[1-G^n]>1/n \Leftrightarrow\\
&\dfrac{G-1}{G-1/G^{n-1}}>1/n\Leftrightarrow\\
&G-G^{1-n}<nG-n.
\end{align}
Then, observe that
\begin{equation}
\dfrac{d^2}{d^2G} (G-G^{1-n})=-n(n-1)G^{-n-1}<0,
\end{equation}
for $n>1$. Thus, 
\begin{equation}
\dfrac{d}{dG} (G-G^{1-n})=1-(1-n)G^{-n}
\end{equation}
is decreasing whereas 
\begin{equation}
\dfrac{d}{dG}(nG-n)=n
\end{equation}
is constant.
Also considering that 
\begin{equation}
\dfrac{d}{dG} (G-G^{1-n})|_{G\rightarrow 1}=n=\dfrac{d}{dG}(nG-n)|_{G\rightarrow 1},
\end{equation}
it follows that
\begin{equation}
\dfrac{d}{dG} (G-G^{1-n})<\dfrac{d}{dG}(nG-n),
\end{equation}
for $n>1$. Since it is also true that as $G\rightarrow 1$, $G-G^{n-1} \rightarrow 0$ and  $nG-n \rightarrow 0$,
\begin{equation}
G-G^{1-n}<nG-n.
\end{equation}
Therefore, if $G>1$, $y_1>1/N$. $\square$
\subsection{Lemma: $1-\alpha^{N-1}N+\alpha^N(N-1)>0$}
For $\alpha=0$, $1-\alpha^{N-1}N+\alpha^N(N-1)=1$. Then, if $\alpha\in (0,1)$ and $N>1$,
\begin{align}
1-\alpha^{N-1}N+\alpha^N(N-1)>&0 \Leftrightarrow\\
[1-\alpha^N]/[N(1-\alpha)]-\alpha^{N-1}>&0 \Leftrightarrow\\
(1/N)(\Sigma_{k=0}^{N-1}\alpha^k)-\alpha^{N-1}>&0 \Leftrightarrow\\
(1/N)(\Sigma_{k=0}^{N-1}\alpha^k)>&(\alpha^{(N-1)N/2})^{\dfrac{2}{N}} \Leftrightarrow\\
(1/N)(\Sigma_{k=0}^{N-1}\alpha^k)>&((\Pi_{k=0}^{N-1}\alpha^k)^{1/N})^2.
\end{align} 
However, since $((\Pi_{k=0}^{N-1}\alpha^k)^{1/N})^2=\alpha^{N-1}<1$,
\begin{equation} 
((\Pi_{k=0}^{N-1}\alpha^k)^{1/N})^2<((\Pi_{k=0}^{N-1}\alpha^k)^{1/N}),
\end{equation}
and since by the AM-GM inequality,
\begin{equation}
(1/N)(\Sigma_{k=0}^{k=N-1}\alpha^k)>((\Pi_{k=0}^{N-1}\alpha^k)^{1/N})
\end{equation}
D.28 must be valid. $\square$.
\subsection{Proof of (22)} 
If $N>1$,
\begin{align}
&N>\dfrac{1-\alpha^{N-1}}{{[1-\alpha^N]/[N(1-\alpha)]-\alpha^{N-1}}} \Leftrightarrow \\
&[1-\alpha^N]/[1-\alpha]-N\alpha^{N-1}>1-\alpha^{N-1} \Leftrightarrow \\
&\Sigma_{k=0}^{N-1}\alpha^k-(N-1)\alpha^{N-1}>1 \Leftrightarrow \\
&(1/(N-1))\Sigma_{k=1}^{N-1}\alpha^k>\alpha^{N-1}.
\end{align}
However, by the AM-GM inequality,
\begin{equation}
(1/(N-1))\Sigma_{k=1}^{N-1}\alpha^k>\sqrt{\alpha^N}.
\end{equation}
D.32 is true if and only if the denominator of D.31 is positive. This is true by the lemma. Also considering that $0<\alpha<1$, and so for $N>2$
\begin{align}
1>&\alpha^{N-2} \Rightarrow \\
\alpha^N>&\alpha^{2N-2} \Rightarrow \\
\sqrt{\alpha^N}>&\alpha^{N-1},
\end{align}
it follows that (D.34) is always true. Since,
\begin{equation} [\alpha/(n-1)][1-\alpha^{N-1}]/[1-\alpha]=\alpha/(n-1)\Sigma_{k=0}^{N-2}\alpha^k>0
\end{equation}
for $\alpha\in (0,1)$ and $N\geq 2$,
\begin{equation} \dfrac{1-\alpha^{N-1}}{{[1-\alpha^N]/[N(1-\alpha)]-\alpha^{N-1}}}> R(\alpha).
\end{equation}
Therefore, $N>R$ for $N>2$. However, if $N=2$, 
\begin{equation}
\dfrac{1-\alpha^{N-1}}{{[1-\alpha^N]/[N(1-\alpha)]-\alpha^{N-1}}}=2.
\end{equation}
Applying (D.40) to (D.41), for $\alpha\in (0,1)$ and $N\geq 2$, $N>R(\alpha)$. $\square$

\subsection{Proof that as $n/N \rightarrow 0$, $R(\alpha)>N(1-\alpha)$}
As $n/N \rightarrow 0$, $R(\alpha)\rightarrow N\dfrac{1-\alpha-\alpha^{N-1}+\alpha^N}{1-\alpha^{N-1}N+\alpha^N(N-1)}$. Hence,
\begin{align}
N(1-\alpha)<&N\dfrac{1-\alpha-\alpha^{N-1}+\alpha^N}{1-\alpha^{N-1}N+\alpha^N(N-1)} \Leftrightarrow\\
(1-\alpha)(1-\alpha^{N-1}N+\alpha^N(N-1))<&1-\alpha-\alpha^{N-1}+\alpha^N \Leftrightarrow\\
1-\alpha^{N-1}N+\alpha^N(N-1)<&1-\alpha^{N-1} \Leftrightarrow\\
\alpha^N(N-1)<&\alpha^{N-1}(N-1) \Leftrightarrow \\
\alpha^N<\alpha^{N-1},
\end{align}
which is true for $\alpha\in(0,1)$. D.43 holds if and only if the denominator of the right-hand-side of D.42 is positive. This is true by the lemma. $\square$

\section{Proof that $R(\alpha)$ Is Strictly Decreasing on $[0,1)$}
Let 
\begin{equation}
F(\alpha)=r([1-\alpha^N]/[N(1-\alpha)]-\alpha^{N-1})-(1-\alpha^{N-1}).
\end{equation}
As shown in Hauert et al. (2002b), $F$ on $(0,1)$ has no root for $r \leq 2$. The preceding result does not hold, though, if $N=2$. We address the case for which $N=2$ at the end of the following proof. For now, we suppose $N>2$. Then, for every $r>2$ there exists exactly one $\alpha$ such that $F=0$. We consider \begin{equation}
Q(\alpha)=\dfrac{1-\alpha^{N-1}}{[1-\alpha^N]/[N(1-\alpha)]-\alpha^{N-1}}.
\end{equation}
Q gives the values of r given $\alpha$ for which F is zero. Hence, Q is injective where it is defined. Since $[1-\alpha^N]/[N(1-\alpha)]-\alpha^{N-1}$ is positive on $[0,1)$ by the lemma in Appendix D, Q is defined and thus injective on (0,1). Thus, Q is either strictly decreasing or strictly increasing on (0,1). However,
\begin{equation}
\lim_{\alpha\rightarrow 0} Q(\alpha) = N,
\end{equation}
and
\begin{equation}
\lim_{\alpha\rightarrow 1} Q(\alpha)=2,
\end{equation}
applying l'Hôpital's rule twice. Since Q is continuous on (0,1) there exist $\delta_1<1/2$ and $\delta_2<1/2$ such that for $\alpha \in (0, 0+\delta_1)$ and for $\alpha \in (1-\delta_2 ,1)$, $|Q(\alpha)-N|<1/3$ and $|Q(\alpha)-2|<1/3$, respectively. Choosing arbitrary $c_1 \in (0,0+\delta_1)$ and $c_2 \in (1-\delta_2,1)$, it follows that for $N>2$, $Q(c_1)>Q(c_2)$ and $c_1<c_2$. Hence, Q must be strictly decreasing on $(0,1)$. Moreover, $Q(0)=N$. Also considering that $Q<N$ on $(0,1)$, as shown in the proof of (22), Q is strictly decreasing on $[0,1)$. Then, we let the numerator of Q be \begin{equation}
S(\alpha)=1-\alpha^{N-1},
\end{equation}
and note that S is strictly decreasing but positive on $[0,1)$. Next, we let the denominator of Q be
\begin{equation}
T(\alpha)=[1-\alpha^N]/[N(1-\alpha)]-\alpha^{N-1},
\end{equation}
which is positive in [0,1) by the lemma in Appendix D. 
Lastly we let
\begin{equation}
U(\alpha)=[\alpha/(n-1)][1-\alpha^{N-1}]/[1-\alpha],
\end{equation}
which is non-negative on [0,1) since it is the product of three non-negative terms. Also, $[1-\alpha^{N-1}]/[1-\alpha]=\Sigma_{k=0}^{N-2}\alpha^k$ for $N\geq2$, a strictly increasing function of $\alpha$ for $\alpha\geq0$ if $N>2$ and constant if $N=2$. Since $\alpha/(n-1)$ is strictly increasing, it follows that $U(\alpha)$ is also strictly increasing on $[0,\infty)$ for $N \geq 2$. 
Next, note that 
\begin{equation}
R(\alpha)=S(\alpha)/[T(\alpha)+U(\alpha)],
\end{equation}
and consider any $\alpha_1, \alpha_2 \in [0,1)$ such that $\alpha_1<\alpha_2$. Then,
\begin{align}
R(\alpha_1)&>R(\alpha_2) \Leftrightarrow \\
S(\alpha_1)/[T(\alpha_1)+U(\alpha_1)]&>S(\alpha_2)/[T(\alpha_2)+U(\alpha_2)] \Leftrightarrow \\
S(\alpha_1)T(\alpha_2)+S(\alpha_1)U(\alpha_2)&>S(\alpha_	2)T(\alpha_1)+S(\alpha_2)U(\alpha_1).
\end{align} 
However, since $S$ is strictly decreasing, $S(\alpha_1)>S(\alpha_2)$. Also considering that since $U$ is strictly increasing, $U(\alpha_2)>U(\alpha_1)$, and that $S$ is positive and $U$ is non-negative with a zero only at $\alpha=0$, it follows that \begin{equation}S(\alpha_1)U(\alpha_2)>
S(\alpha_2)U(\alpha_1).
\end{equation}
Furthermore, since Q is strictly decreasing and T is positive, 
\begin{equation}
S(\alpha_1)T(\alpha_2)>S(\alpha_2)T(\alpha_1).
\end{equation}
(E.12) and (E.13) together imply that E.11 is valid. Thus, R is strictly decreasing on $[0,1)$ for $N>2$. However, if $N=2$, then the only change from the above proof is that Q is constant rather than strictly decreasing. Then, (E.12) still holds, and we replace (E.13) by
\begin{equation}
S(\alpha_1)T(\alpha_2)=S(\alpha_2)T(\alpha_1).
\end{equation}
Thus, (E.11) still holds. Hence, R is strictly decreasing on $[0,1)$ for $N\geq 2$. $\square$
\section{Derivation of $\pi_1$ (equation 24)} 
The payoff matrix for a two person public goods game in which cooperators invest 1 unit which is then multiplied by r and distributed equally among all players is 
\begin{equation*}
\begin{array}{c|cc}
 & c & d \\
 \hline
c& r-1 & r/2-1 \\
d& r/2 & 0
\end{array}
\end{equation*}
Then, we suppose that there are i players of type 1 in a population of n individuals, and that the remaining individuals are of type 2. We let a player of type 1 be one of the players invited to play in the two person PGG and call that player "player A". Next, we let $A_c$, $A_d$, $A_n$, and $A_n^c$ represent the events where player A cooperates, defects, does not participate, and participates, respectively. We suppose "player B" is the other individual invited to play. We let $B_c$, $B_d$, $B_n$ be the events where player B cooperates, defects, and does not participate, respectively. Lastly, we let $E_1$ and $E_2$ be the events where Player B is of type 1 and of type 2, respectively. Denoting the intersection of any two events F and G by FG, and the probability that an event F occurs by p(F),
\begin{align*}
\begin{split}
\pi_1=&(r-1)[p(A_cE_1B_c)+p(A_cE_2B_c)]+(r/2-1)[p(A_cE_1B_d)+p(A_cE_2B_d)]\\&+r/2[p(A_dE_1B_c)+p(A_dE_2B_c)]+\sigma[p(A_n)+p(A_n^cE_1B_n)+p(A_n^cE_2B_n)]\\
=&(r-1)[p(A_c)p(E_1|A_c)p(B_c|A_cE_1)+p(A_c)p(E_2|A_c)p(B_c|A_cE_2)]\\&+(r/2-1)[p(A_c)p(E_1|A_c)p(B_d|A_cE_1)+p(A_c)p(E_2|A_c)p(B_d|A_cE_2)]\\&+r/2[p(A_d)p(E_1|A_d)p(B_c|A_dE_1)+p(A_d)p(E_2|A_d)p(B_c|A_dE_2)]\\&+\sigma[p(A_n)+p(A_n^c)p(E_1|A_n^c)p(B_n|A_n^cE_1)+p(A_n^c)p(E_2|A_n^c)p(B_n|A_n^cE_2)]\\
=&(r-1)\beta_1(1-\alpha_1)[p(E_1|A_c)\beta_1(1-\alpha_1)+p(E_2|A_c)\beta_2(1-\alpha_2)]\\&+(r/2-1)\beta_1(1-\alpha_1)[p(E_1|A_c)(1-\beta_1)(1-\alpha_1)+p(E_2|A_c)(1-\beta_2)(1-\alpha_2)]\\&+r/2(1-\beta_1)(1-\alpha_1)[p(E_1|A_d)\beta_1(1-\alpha_1)+p(E_2|A_d)\beta_2(1-\alpha_2)]\\&+\sigma[\alpha_1+(1-\alpha_1)[p(E_1|A_n^c)\alpha_1+p(E_2|A_n^c)\alpha_2]\\
=&(r-1)\beta_1(1-\alpha_1)[\dfrac{i-1}{n-1}\beta_1(1-\alpha_1)+\dfrac{n-i}{n-1}\beta_2(1-\alpha_2)]\\&+(r/2-1)\beta_1(1-\alpha_1)[\dfrac{i-1}{n-1}(1-\beta_1)(1-\alpha_1)+\dfrac{n-i}{n-1}(1-\beta_2)(1-\alpha_2)]\\&+r/2(1-\beta_1)(1-\alpha_1)[\dfrac{i-1}{n-1}\beta_1(1-\alpha_1)+\dfrac{n-i}{n-1}\beta_2(1-\alpha_2)]\\&+\sigma[\alpha_1+(1-\alpha_1)[\dfrac{i-1}{n-1}\alpha_1+\dfrac{n-i}{n-1}\alpha_2]\\
=&\dfrac{n - i}{n - 1}(1 - \alpha_1)(1 - \alpha_2)(r\beta_2/2 + r\beta_1/2- \beta_1) + \dfrac{i - 1}{n - 1}(1 - \alpha_1)^2\beta_1(r - 1)\\& + \sigma\alpha_1 + \sigma(1 - \alpha_1)(\dfrac{n - i}{n - 1}\alpha_2 + \dfrac{i - 1}{n - 1}\alpha_1)
\end{split}
\end{align*} 
\section{Justification of Approximations}
\subsection{Approximation for $R(\alpha)$ as $N\rightarrow \infty$, $\dfrac{N}{n-1}=c$}
We let $c$ be a real number in [0,1]. As $N\rightarrow\infty$, $\alpha^{N-1}N$, $\alpha^N(N-1-N/(n-1))$, $\alpha^{N-1}$, and $\alpha^N$ $\rightarrow 0$ as long as $\alpha\nrightarrow 1$. Hence, for $\alpha\nrightarrow 1$, 
\begin{align*} 
R(\alpha)=&N\dfrac{1-\alpha-\alpha^{N-1}+\alpha^N}{1+\alpha N/(n-1)-\alpha^{N-1}N+\alpha^N(N-1-N/(n-1))}\\
\approx & N(1-\alpha)/(1+c\alpha).
\end{align*}
However, applying l'Hôpital's rule yields  $\lim_{\alpha \rightarrow 1}R(\alpha)=0$, which is $\lim_{\alpha\rightarrow 1}N(1-\alpha)/(1+1/2\alpha)$. $\square$
\subsection{Approximation for $R(\alpha)$ for $n>>N>>0$}
As $n \rightarrow \infty$, $N \rightarrow \infty$, $N/n \rightarrow 0$, for $alpha \nrightarrow 1$,
\begin{align*} 
R(\alpha)=&N\dfrac{1-\alpha-\alpha^{N-1}+\alpha^N}{1+\alpha N/(n-1)-\alpha^{N-1}N+\alpha^N(N-1-N/(n-1))}\\
\approx & N(1-\alpha).
\end{align*}
However, as in the preceding proof, applying l'Hôpital's rule yields  $\lim_{\alpha \rightarrow 1}R(\alpha)=0$, which is $\lim_{\alpha\rightarrow 1}N(1-\alpha)$. $\square$
\\\\


\textbf{References}
\begin{thebibliography}{00}
\bibitem[Antal et al. (2009)]{01}
Antal, T., Ohtsuki, H., Wakeley, J., Taylor, P.D. and Nowak, M.A., 2009. Evolution of cooperation by phenotypic similarity. Proceedings of the National Academy of Sciences, 106(21), pp.8597-8600.
\bibitem[Axelrod (1984)]{02}
Axelrod, R., 1984. The Evolution of Cooperation. Basic books. 

\bibitem[Battiston et al.(2017)]{Battiston_NJP17}
Battiston, F., Perc, M. and Latora, V., 2017. Determinants of public cooperation in multiplex networks. New Journal of Physics.

\bibitem[Boyd et al. (2010)]{03}
Boyd, R., Gintis, H. and Bowles, S., 2010. Coordinated punishment of defectors sustains cooperation and can proliferate when rare. Science, 328(5978), pp.617-620.
\bibitem[Hauert et al. (2002a)]{04}
Hauert, C., De Monte, S., Hofbauer, J. and Sigmund, K., 2002. Volunteering as red queen mechanism for cooperation in public goods games. Science, 296(5570), pp.1129-1132.
\bibitem[Hauert et al. (2002b)]{05}
Hauert, C., De Monte, S., Hofbauer, J. and Sigmund, K., 2002. Replicator dynamics for optional public good games. Journal of Theoretical Biology, 218(2), pp.187-194.
\bibitem[Hauert et al (2008)]{06}
Hauert, C., Traulsen, A., née Brandt, H.D.S., Nowak, M.A. and Sigmund, K., 2008. Public goods with punishment and abstaining in finite and infinite populations. Biological theory, 3(2), pp.114-122.
\bibitem[Hölldobler and Wilson (2009)]{07}
Hölldobler, B. and Wilson, E.O., 2009. The Superorganism: The Beauty, Elegance, and Strangeness of Insect Societies. W. W. Norton $\&$ Company.
\bibitem[Imhof and Nowak (2010)]{08}
Imhof, L.A. and Nowak, M.A., 2010. Stochastic evolutionary dynamics of direct reciprocity. Proceedings of the Royal Society of London B: Biological Sciences, 277(1680), pp.463-468.
\bibitem[Nadell et al. (2008)]{09}
Nadell, C.D., Xavier, J.B., Levin, S.A. and Foster, K.R., 2008. The evolution of quorum sensing in bacterial biofilms. PLoS biology, 6(1), p.e14.
\bibitem[Nowak (2006a)]{10}
Nowak, M.A., 2006. Evolutionary Dynamics: Exploring the Equations of Life. Cambridge, Mass, Belknap Press.
\bibitem[Nowak (2006b)]{11}
Nowak, M.A., 2006. Five rules for the evolution of cooperation. science, 314(5805), pp.1560-1563.
\bibitem[Pacheco et al. (2015)]{12}
Pacheco, J.M., Vasconcelos, V.V., Santos, F.C. and Skyrms, B., 2015. Co-evolutionary dynamics of collective action with signaling for a quorum. PLoS computational biology, 11(2), p.e1004101.
\bibitem[Priklopil et al. (2017)]{13}
Priklopil, T., Chatterjee, K. and Nowak, M., 2017. Optional interactions and suspicious behaviour facilitates trustful cooperation in prisoners dilemma. Journal of Theoretical Biology.
\bibitem[Schoenmakers et al. (2014)]{13}
Schoenmakers, S., Hilbe, C., Blasius, B. and Traulsen, A., 2014. Sanctions as honest signals–The evolution of pool punishment by public sanctioning institutions. Journal of theoretical biology, 356, pp.36-46.

\bibitem[Szolnoki \& Perc(2015a)]{Szolnoki_RSI15}
Szolnoki, A. and Perc, M., 2015. Conformity enhances network reciprocity in evolutionary social dilemmas. Journal of The Royal Society Interface, 12(103), p.20141299.


\bibitem[Szolnoki \& Perc(2015b)]{Szolnoki_PRSB15}
Szolnoki, A. and Perc, M., 2015, October. Antisocial pool rewarding does not deter public cooperation. In Proc. R. Soc. B (Vol. 282, No. 1816, p. 20151975). The Royal Society.

\bibitem[Traulsen et al. (2009)]{14}
Traulsen, A., Hauert, C., De Silva, H., Nowak, M.A. and Sigmund, K., 2009. Exploration dynamics in evolutionary games. Proceedings of the National Academy of Sciences, 106(3), pp.709-712.
\bibitem[Traulsen and Nowak (2006)]{15}
Traulsen, A. and Nowak, M.A., 2006. Evolution of cooperation by multilevel selection. Proceedings of the National Academy of Sciences, 103(29), pp.10952-10955.
\bibitem[Traulsen et al. (2007)]{16}
Traulsen, A., Pacheco, J.M. and Nowak, M.A., 2007. Pairwise comparison and selection temperature in evolutionary game dynamics. Journal of theoretical biology, 246(3), pp.522-529.
\bibitem[Trivers (1971)]{17}
Trivers, R.L., 1971. The evolution of reciprocal altruism. The Quarterly review of biology, 46(1), pp.35-57.
\end{thebibliography}
\end{document}